\documentclass[amsmath,amssymb,aps,pre,twocolumn,showkeys,showpacs,superscriptaddress,floatfix]{revtex4-1}
\usepackage{amsopn}
\usepackage{amsfonts}
\usepackage{amssymb}
\usepackage{amsmath}
\usepackage{bm}
\usepackage{siunitx}
\usepackage[pdftex]{graphicx}

\DeclareMathOperator*{\argmin}{arg\,min}

\begin{document}
\title{Real-space analysis of scanning tunneling microscopy 
topography datasets using sparse modeling approach }
\date{\today}
\author{Masamichi J. \surname{Miyama}}
\affiliation{Department of Pure and Applied Sciences, University of Tokyo, Komaba, Meguro-ku, Tokyo 153-8902, Japan}
\author{Koji \surname{Hukushima}}
\affiliation{Department of Pure and Applied Sciences, University of Tokyo, Komaba, Meguro-ku, Tokyo 153-8902, Japan}
\affiliation{Center for Materials Research by Information Integration,
National Institute for Materials Science, 1-2-1 Sengen, Tsukuba, 
Ibaraki 305-0047, Japan}

\begin{abstract}
  A sparse modeling approach is proposed for analyzing scanning
 tunneling microscopy topography data, which contains numerous peaks
 corresponding to surface atoms. The method, based on the relevance
 vector machine with $\text{L}_1$ regularization and $k$-means clustering,
 enables separation of the peaks and atomic center positioning with
 accuracy beyond the resolution of the measurement grid.
 The validity and efficiency of the proposed method are demonstrated
 using synthetic data in comparison to the conventional least-square method.
 An application of the proposed method to experimental data of a metallic oxide thin film clearly indicates the existence of defects and corresponding local lattice deformations. 
 \end{abstract}

\maketitle

\section{Introduction}
Scanning tunneling microscopy (STM) is an experimental technique that
enables observation of a material surface at atomic-scale
resolution \cite{STM1, STM2, TersoffHamann}. An electron-density topography map
is obtained with STM by measuring the tunneling current between the
surface to be observed and an atomic-scale conducting tip with an
applied bias voltage.
Since the invention of STM, various types of scanning probe
microscopies, such as atomic force microscope, have been developed and
used for measuring surface topography and physical properties of
materials surfaces. 

Several interesting phenomena on the surfaces have been shown to be caused by local
strain induced by impurities and/or defects.
For example, the critical
temperature of High-$T_{\text{c}}$ cuprate superconductors significantly
depends on local strain \cite{YOkada, Saini, Deutscher, Zeljkovic}.  
Fourier transforms are often used for extracting certain properties of surface structures, such as the set of lattice vectors of a surface reconstruction structure \cite{Gai}. For a clean crystalline surface structure, Fourier transforms can be used to accurately estimate the atomic positions and associated local strain from the perfect lattice structure. However, thin films of metallic oxides are generally not clean surface structures, and it is difficult to extract local structural information from STM topography data. In fact, the desired local information for thin film structures can be obscured behind noise in the Fourier transform. 
Hence, a new methodology for performing real-space data analysis beyond the Fourier transform is highly desired. 

In this study, we propose a data-analysis methodology for extracting the
atomic arrangement from noisy STM topography. Our method is based on the
fact that the STM topography data
for a given surface can be represented by the superposition of 
suitable basis functions with noise,
each of which is a spatially localized
with a center corresponding to the
location of atom. 
The basis function is characterized by 
the set of parameters,
which includes the center position, amplitude, and shape of the basis
function.

Our strategy decomposes 
a given STM data set into the basis functions
with determining the set of parameters in the
data model. This strategy could be accomplished by
using the least-squares method. In fact, when
the number of the peaks $N_{\text{peak}}$ and a
shape parameter of the basis function are 
known in advance, this simple strategy is
effective. However, because the typical number
of atoms assumed here could be more than ten
thousand and the associated number of data
points could be more than one million, it is
difficult to know $N_{\text{peak}}$
beforehand. Also, the shape parameters of the
basis function are unknown a priori in general.  

To establish a methodology for analyzing STM topography with an
unspecified number of atoms, we use a relevance vector machine (RVM)
\cite{RVM} as the data model and a maximum a posteriori (MAP)
estimation, which is based on the framework of Bayesian inference, to
determine the model parameters. As the prior distribution in MAP
estimation, we introduce a Laplace prior which is equivalent to 
the least absolute shrinkage and selection operator (LASSO)
regression \cite{LASSO}. Depending on the measurement resolution, the
number of data points is typically much larger than that of atoms, so
the variables that we 
extract from the data can be ``sparse.'' Using LASSO permits model
inference with emphasis on the sparsity of the data.
Recently, sparse modeling has been applied to a wide range of problems dealing with
high-dimensional data.  Our proposed  method is regarded as a sparse
modeling for STM topography data analysis. 

In this paper, we present the procedures used for extracting the atomic
positions and peak amplitudes included in the STM topography images; we
discuss not only the method for determining the model parameters but
also the method for validating the models. First, we apply our method to
synthesis data, and we examine the accuracy of our estimation. Then, we
report the results of our model application to actual STM topography
data obtained with a metallic oxide thin film.

\section{Model and Method}
\begin{figure}
  \includegraphics[width=.41\linewidth,clip]{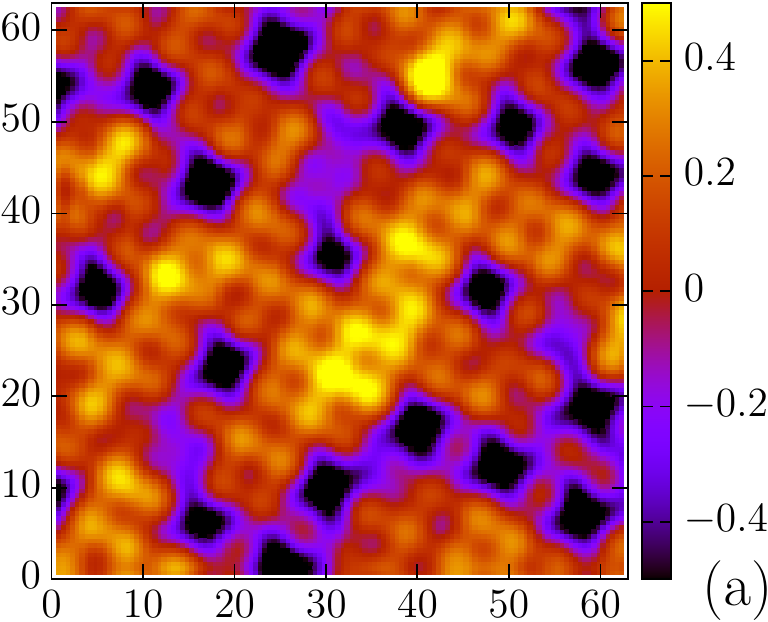}
  \includegraphics[width=.575\linewidth,clip]{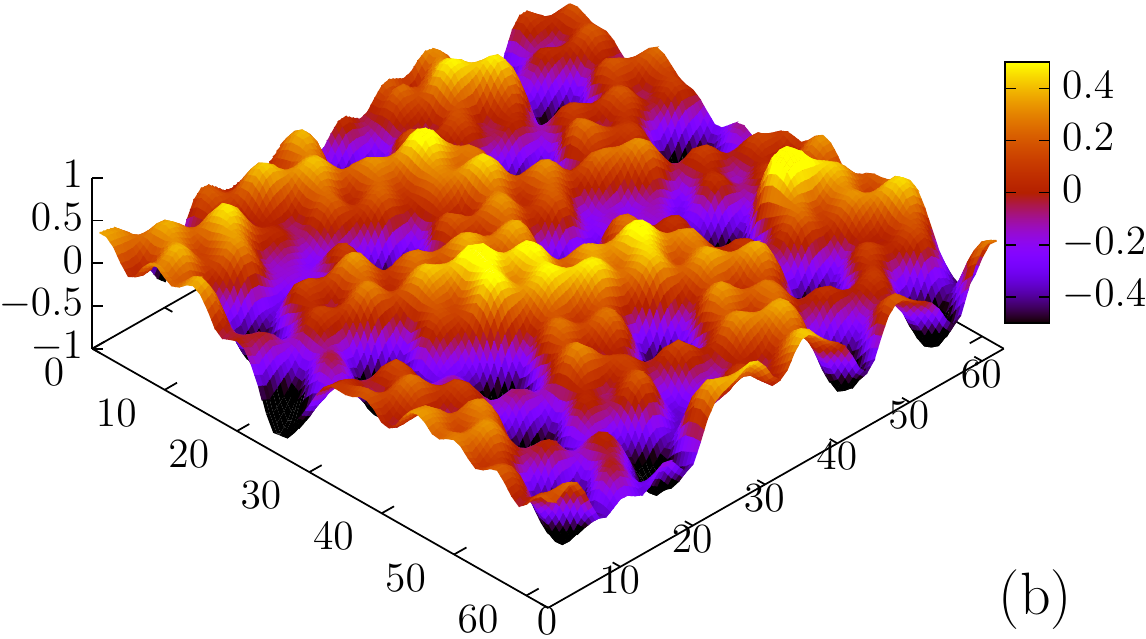}
  \caption{A typical STM topography image of a $\text{SrVO}_3$ thin
 film: (a) top view 
 and (b) bird's-eye view. Each peak corresponds to an atom, and the dark
 spots are the atomic (oxygen) defects. The original data was provided
 by Y.~Okada and T.~Hitosugi. }
  \label{fig:1}
\end{figure}
\subsection{Data model}
Typical topography data obtained by STM measurements of $\text{SrVO}_3$
is shown in Fig.~\ref{fig:1}. The STM topography picture typically shown
in literature is the top view shown on the left of Fig.~\ref{fig:1}. The
surface of $\text{SrVO}_3$ is relatively clean and flat in comparison to
the surfaces of other metallic oxide compounds. Nevertheless, it is
noticed from the bird's-eye view (Fig.~\ref{fig:1} (b)) that the STM
image has a rugged structure. This structure may be due to atomic-scale
fluctuation or the STM tip condition. Each peak in the figure is
considered to correspond to an atom, and dark spots often 
indicate the existence of atomic defects.
Note that STM is generally responsible for indicating the electronic
state underlying the surface, not the atom itself.  
Our aim in this work is to decompose such STM topography data into the
peaks, assumed to be resulting from each atom.  

This process is formally similar to the peak decomposition of spectral
data measured in various natural science experiments. Recently, a
statistical analysis technique based on Bayesian inference was used to
successfully extract a finite number of peaks for a one-dimensional data
spectrum with noise \cite{Nagata, Igarashi}. In this sense, our problem
might be considered as peak decomposition in a two-dimensional (2D) data spectrum. The number of peaks in this study, however, is much larger than those attained in the previous study. Thus, a different numerical calculation strategy is required for treating the large set of data.

In this study, our framework is based on the RVM . The pixel data is denoted as $\bm{y}=(y_1,\cdots,y_D)$ with $D$ being the total number of pixels, and the vector $\bm{x}$ represents the weight of the STM source signal to be estimated. The weight $x_i$ is defined on an artificial array point, which is generally different from the original pixel array for $\bm{y}$. The dimension of the vector $\bm{x}$, which is the number of array points introduced, is denoted by $N$. The number $N$ can be chosen independently of $D$ depending on the resolution of the estimation. Assuming an explicit functional form of the measurement matrix, which we discuss later, our task is reduced to inferring relevant components in vector $\bm{x}$ for a given vector $\bm{y}$.

Using a $D \times N$ measurement matrix $\hat{A}$, our data model is expressed as 
\begin{equation}
 \bm{y} = \hat{A}\bm{x}+\bm{\epsilon},
\label{eqn:Dmodel}
\end{equation}
where $\bm{\epsilon}$ is a noise vector with dimension $D$ associated with the observation. For simplicity, each element of the vector $\bm{\epsilon}$ is assumed to be \textit{iid} Gaussian random variables with zero mean and variance $\sigma_\epsilon$: 
\begin{equation}
  P(\bm{\epsilon}) = \prod_{d=1}^D \frac{1}{\sqrt{2 \pi \sigma_{\epsilon}^2}}
  \exp \left( -\frac{\epsilon_d^2}{2\sigma_{\epsilon}^2} \right).
  \label{eqn:GaussianNoise}
\end{equation}
In other words, the noise property is independent of the pixel position, and coherent noise such as that induced by the STM tip or by the surface condition is not considered. 

In our method, the vector $\bm{x}$ is estimated by a posterior distribution $P(\bm{x}|\bm{y})$ for a given pixel data $\bm{y}$. With Bayes' theorem, the a posterior distribution is expressed as 
\begin{equation}
  P(\bm{x}|\bm{y}) = \frac{P(\bm{y}|\bm{x}) P(\bm{x})}
  {\sum_{\bm{x}} P(\bm{y}|\bm{x}) P(\bm{x})},
  \label{eq:Bayes}
\end{equation}
where $P(\bm{y}|\bm{x})$ and $P(\bm{x})$ are the likelihood function and
a prior distribution, respectively. We employ the MAP estimation in
which the value of $\bm{x}$ is chosen by maximizing the posterior
distribution. The likelihood function is given by the noise distribution
of Eq.~(\ref{eqn:GaussianNoise}) as 
\begin{equation}
  P(\bm{y}|\bm{x}) = P(\bm{\epsilon})
  = \prod_{d=1}^D \frac{1}{\sqrt{2 \pi \sigma_{\epsilon}^2}}
  \exp \left(-\frac{\lVert\bm{y}-\hat{A}\bm{x} \rVert_2^2}
  {2\sigma_{\epsilon}^2}\right),
\end{equation}
where $\Vert\cdots\rVert_2$ denotes the L$_2$ norm. The prior distribution in Eq.~(\ref{eq:Bayes}) used here is given by a Laplace prior over $\bm{x}$:
\begin{equation}
  P(\bm{x}) \propto \exp \left(-\lambda \lvert \bm{x} \rvert_1 \right), 
\end{equation}
where $\lambda$ is a hyperparameter and $\lvert\cdots\rvert_1$ is the
L$_1$ norm. The prior distribution usually reduces the number of
non-zero elements of the vector $\bm{x}$. We assume sparsity of the
vector $\bm{x}$ based on the reasonable assumption that the number of
signal sources from existing atoms is significantly smaller than that of
the pixel arrays.
The present approach is called the sparse modeling. 
It is emphasized that our framework does not specify the number of peaks
$N_{\rm peak}$ at the present stage. All the element of $\bm{x}$ could
be the peak centers in principle and the sparse modeling is used for a
sparse solution for $\bm{x}$ with a small number of non-zero elements

\subsection{Measurement matrix and MAP estimate}
Our data model of Eq.~(\ref{eqn:Dmodel}), represented by a linear
relation with additive noise, means that the observation vector $\bm{y}$
is a superposition of the 
basis functions and that the relevance vector $\bm{x}$ is a weight factor. In this work, we assume the kernel function is an isotropic 2D Gaussian function in which the element of the measurement matrix in Eq.~(\ref{eqn:Dmodel}) is given by
\begin{equation}
  A_{di}(r_{di};\sigma) =
  \frac{1}{\sqrt{2\pi\sigma}}\exp\left(-\frac{r_{di}^2}{2\sigma^2}\right),
  \label{eqn:Mmatrix}
\end{equation}
where $A_{di}$ is an element of the measurement matrix $\hat{A}$, $\sigma$ represents the variance, and $r_{di}$ is the spatial distance between the position of measurement $y_d$ and that of signal source $x_i$.

Note that there is no theoretical or physical basis for choosing  the 2D 
Gaussian function. In Ref.~\onlinecite{Gai}, the 2D isotropic Gaussian
function with the covariance matrix $\Sigma=\sigma \bm{I}$, where
$\bm{I}$ is an identity matrix, is used to fit peaks in STM topography
data. In the field of optics, the width of the point spread function,
which corresponds to our basis function, can be measured by an
independent experiment a priori. In that case, an algorithm based on
the maximum-likelihood method works well \cite{Ashida}. 
However, it is difficult to know the value of $\sigma$ from a
calibration experiment in STM because the target surfaces as well as the
tip states are sensitive to experimental conditions.
Our problem 
is more difficult than a peak decomposition problem with known $\sigma$
in the sense that simultaneous inference of peaks and the value of
$\sigma$ is to be solved from the input data $\bm{y}$. 

The MAP estimate with respect to $\bm{x}$ is equivalent to the minimization of the cost function $E(\bm{x};\bm{y},\lambda, \bm{\mu})$,
\begin{equation}
  E(\bm{x}; \bm{y},\lambda, \bm{\mu}) = \frac{1}{2\sigma_\epsilon^2}
  \lVert \bm{y}-\hat{A}\bm{x} \rVert_2^2 + \lambda \lvert \bm{x} \rvert_1,
  \label{eqn:CostFunc}
\end{equation}
where $\bm{\mu}$ denotes a set of unknown parameters in the measurement
matrix $\hat{A}$. The inference scheme with the prior distribution is
known as the least absolute shrinkage and selection operator (LASSO)
\cite{LASSO}, and the hyperparameter $\lambda$ determines the strength
of the sparsity. Our inference scheme is the vector machine with
$\text{L}_1$ regularization, which is equivalent the so-called L1VM
\cite{L1VM}. In our case, the parameter $\bm{\mu}$ includes the variance
$\sigma$ in Eq.~(\ref{eqn:Mmatrix}). Without loss of generality, the
unit of the cost function is set to $\sigma_\epsilon^{-2}$, and thus the
cost function is represented as a function of $\bm{x}, \lambda$,  and
$\bm{\mu}$. This resulting problem is an optimization problem. In this
work, we use a fast iterative shrinkage-thresholding algorithm (FISTA)
\cite{FISTA} for minimizing the cost function, which is popularly used
in $\text{L}_1$ optimization problems. For $\lambda = 0$, minimization
of the cost function is reduced to the least-squares method, and for a
sufficiently large value of $\lambda$, the trivial solution $\bm{x} =
\bm{0}$ is obtained. Therefore, an appropriate value of $\lambda$ is
expected to exist between these two extremes, and $\lambda$ can be
determined  as a consequence of the competition between the data fit and
the sparsity of $\bm{x}$. Unfortunately, an appropriate value of
$\lambda$ is not known a priori. It would be suitable to choose
the value of $\lambda$ to reduce prediction error; however, the
prediction error is difficult to estimate. Instead, a promising method
for determining the hyperparameter $\lambda$, as well as unknown
parameters in the likelihood function, is through cross validation (CV).

\subsection{Cross validation and hyperparameter selection}
In $K$-fold CV, the data set of $\bm{y}$ is divided into $K$ subsets, which are denoted by $\{\bm{y}^{(k)}\} = \{y_{\Lambda^{(k)}_1}, \dots, y_{\Lambda^{(k)}_{D/K}}\}$ with $k = 1, \dots, K$. Here, $\Lambda^{(k)}$ is an index set of the elements contained in $k$-th subset. The subsets are chosen randomly from the original $\bm{y}$, and each element of $\bm{y}$ appears once in the subsets. Using the data set $\overline{\bm{y}}^{(k)} = \bm{y}\setminus \bm{y}^{(k)}$ as a training set, we obtain the optimal solution $\bm{x}^{(k)}$ that minimizes the cost function $E(\bm{x}; \overline{\bm{y}}^{(k)}, \lambda, \bm{\mu})$. Then, for each test set $\bm{y}^{(k)}$, we calculate the CV error as
\begin{equation}
  L^{(k)}(\lambda,\bm{\mu})
  = \frac{1}{2} \lVert \bm{y}^{(k)}-\hat{A}^{(k)} \bm{x}^{(k)} \rVert_2^2,
\end{equation}
where the measurement matrix $\hat{A}^{(k)}$ for the partial data set $\bm{y}^{(k)}$ is given by $\hat{A}^{(k)} = (\bm{A}_{\Lambda^{(k)}_1}, \dots, \bm{A}_{\Lambda^{(k)}_{D/K}})^{\text{T}}$ with $\bm{A}_{\Lambda^{(k)}_i}=(A_{\Lambda^{(k)}_i 1}, \dots, A_{\Lambda^{(k)}_i N})$. Averaging over the possible test data set, the averaged CV error is defined by 
\begin{equation}
 \overline{L}^{K}(\lambda,\bm{\mu}) = \frac{1}{K}\sum_{k=1}^KL^{(k)}(\lambda,\bm{\mu}). 
\end{equation}
Regarding the CV error as an estimate of the prediction error, the hyperparameter and unknown parameter are determined by minimizing the averaged CV error. The CV error $\overline{L}^{K}$ is known to equal the true prediction error in the large $K$ limit, so ideally, we should choose a sufficiently large number of $K$. In particular, the case of $K=D$ corresponds to so-called leave-one-out cross validation (LOOCV), which requires $D$ minimization calculations for a given set of $\lambda$ and $\bm{\mu}$. This CV is time-consuming with increasing the data size $D$.

Recently, Obuchi and Kabashima \cite{ObuchiKabashima} have proposed a simplified method for performing LOOCV. Once the minimization of the cost function is computed for the total data set, the LOOCV error $\overline{L}^{\text{LOO}}$ is estimated by the approximated formula given by
\begin{equation}
  \overline{L}^{\text{LOO}}(\lambda, \bm{\mu})
  = \left(\frac{N}{N_0(\epsilon^{\text{th}})}\right)^2
  \sum_{d=1}^D \left(y_d - \sum_{i=1}^N A_{di}x_i \right)^2,
  \label{eq:LOOCV}
\end{equation}
where $N_0(\epsilon^{\text{th}})$ is the number of elements of $\bm{x}$ below the threshold $\epsilon^{\text{th}}$. The value of $\epsilon^{\text{th}}$ may depend on the solver used for minimizing the cost function Eq.~(\ref{eqn:CostFunc}). Using FISTA, $\epsilon^{\text{th}}$ is unambiguously obtained by $\epsilon_{\text{th}} = \lambda/L$, where $L$ is a Lipschitz  constant of the cost function (see Appendix \ref{sec:FISTA}). 

We performed the $K$-fold CV procedure for typical STM data with changing $K$ and confirmed that $\overline{L}^{K}(\lambda, \bm{\mu})$ is almost independent of $K$ for $K \geq 10$. Thus, in the following sections, we present the results of both 10-fold CV and the approximated LOOCV for comparison. 

In our data model, two parameters are to be determined by CV: the LASSO tuning parameter $\lambda$ and the variance of the Gaussian function $\bm{\mu}=\{\sigma\}$. We first determine $\lambda$ for a fixed value of $\sigma$ according to the one-standard-error rule \cite{onese} often used in a LASSO analysis, that is,
\begin{equation}
  \lambda^*(\sigma) = \max_{\lambda}
  \left\{
    \lambda \left|~\Vert\overline{L}^{K}(\lambda)
    - \overline{L}^{K}(\hat{\lambda})\Vert_2
    < \text{SE}(L^{(k)}(\hat{\lambda}))\right.
  \right\},
\end{equation}
where $\hat{\lambda}$ is given by $\hat{\lambda} = \argmin_{\lambda}L^{K}(\lambda)$ and $\text{SE}(\cdots)$ is the standard error of the $K$-fold CV error. After choosing $\lambda^*$ as a function of $\sigma$, we choose a suitable $\sigma$ as the minimizer of the CV error, that is,
\begin{equation}
  \sigma^*=\argmin_{\sigma} \overline{L}^{K} (\lambda^*(\sigma), \sigma).
\end{equation}

  \subsection{Estimation of the peak position }
Our goal is to determine the position of atoms with a reasonable resolution in order to quantify any local distortion of the position. The non-zero elements of the estimated vector $\bm{x}$ will lead to the central peak position. The resolution of each position is, however, limited to the grid size of our data model. Some non-zero elements of the optimized value $x_i$ are localized, and they are separated from each other. Therefore, we can extract the center of peaks from $\bm{x}$ with higher resolution than those obtained with the L1VM grid size using the $k$-means clustering method.  

We suppose that the number of the peaks $N_{\text{peak}}$ in $k$-means clustering is countable for the estimated vector $\bm{x}$. This assumption is based on the fact that the non-zero elements of $\bm{x}$ are highly localized in the L1VM grid space. The center of $k$th peak $\bm{r}_k$ with $k=1, \dots, N_{\text{peak}}$ is initially chosen by a certain pixel $i$ at which the element $x_i$ takes a maximum value within the radius $R$ around pixel $i$. The value of $R$ is appropriately set as a mean distance of the localized elements.

Then, an attributed variable $z_i$ is allocated for each pixel $i$ as 
\begin{equation}
    z_i = \argmin_k d(\bm{i}, \bm{r}_k),
  \label{eq:K-means1}
\end{equation}
where $\bm{i}$ is the position vector at pixel $i$ on the L1VM grid as $\bm{i}=(i_x, i_y)$ and $d(\bm{i}, \bm{r}_k)$ denotes the Euclid distance between the pixel $i$ and the center $\bm{r}_k$. Using the attributed variables, the center is defined by 
\begin{equation}
  \bm{r}_k = \frac{\displaystyle \sum_i \delta_{k, z_i} \theta(x_i) x_i
    \bm{i}}{\displaystyle \sum_i \delta_{k,z_i} \theta(x_i) x_i},
  \label{eq:K-means}
\end{equation}
where $\theta(x)$ is a Heaviside step function, meaning that an element
with a negative value is not considered in  this analysis. Here,
$\bm{r}_k$ 
is a weighted average of the pixel positions when the amplitudes $\bm{x}$ are regarded as the weights. Solving Eq.~(\ref{eq:K-means1}) and (\ref{eq:K-means}) iteratively, we obtain the centers of clusters $\bm{r}^*_k$.

\section{Numerical Results}
\subsection{Synthetic data and typical examples of estimated data}
First, we examine the validity and reliability of the proposed method using a synthetic data set. The synthetic data are generated by the following procedure. For the given primitive basis vectors, $\bm{a}_1$ and $\bm{a}_2$ of the 2D lattice, the lattice vector $\bm{r}_i$ of $i$-th lattice point is defied as
\begin{equation}
 \bm{r}_i = m\bm{a}_1+n\bm{a}_2+\bm{\xi}_i,
\end{equation}
where $m$ and $n$ are integers and $\bm{\xi}_i$ is a uniformly random vector representing  a local lattice distortion. All atoms are allocated at the lattice points in the region $[0{}\colon{}\ell] \times [0{}\colon{}\ell]$, where the length unit of the lattice data is set to $1 \text{px}$. Some lattice points are attributed to vacancy sites, which are randomly chosen with the probability $\rho_{\text{vac}}$. The number of peaks $N_{\text{peak}}$ is given by $N_{\text{peak}} = (1 - \rho_{\text{vac}}) N_{\text{tot}}$ with $N_{\text{tot}}$ being the number of lattice points in the region under consideration. Thus, the atom positions to be inferred from the imaging data are determined as $\{\hat{\bm{r}}_k\} \text{ ($k = 1, \dots, N_{\text{peak}}$)}$.

\begin{figure}[ht]
  \includegraphics[width=0.4125\linewidth,clip]{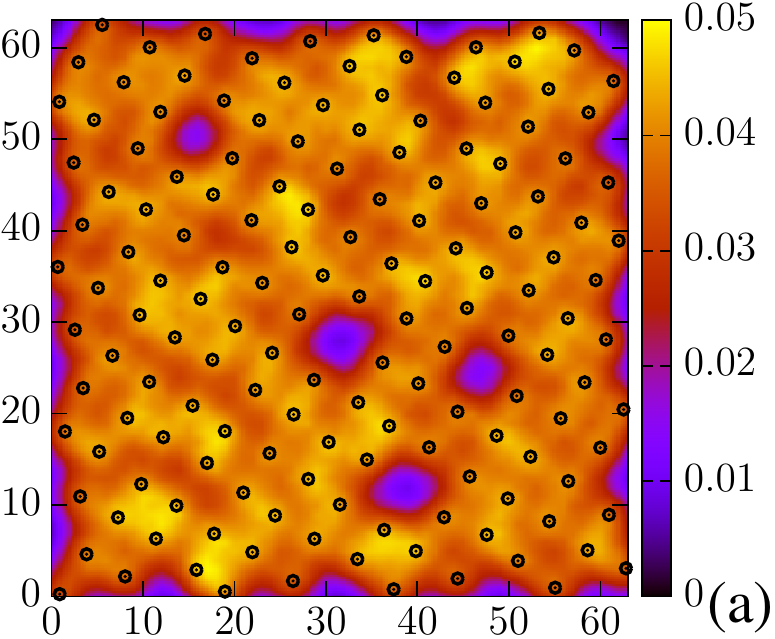}
  \includegraphics[width=0.5725\linewidth,clip]{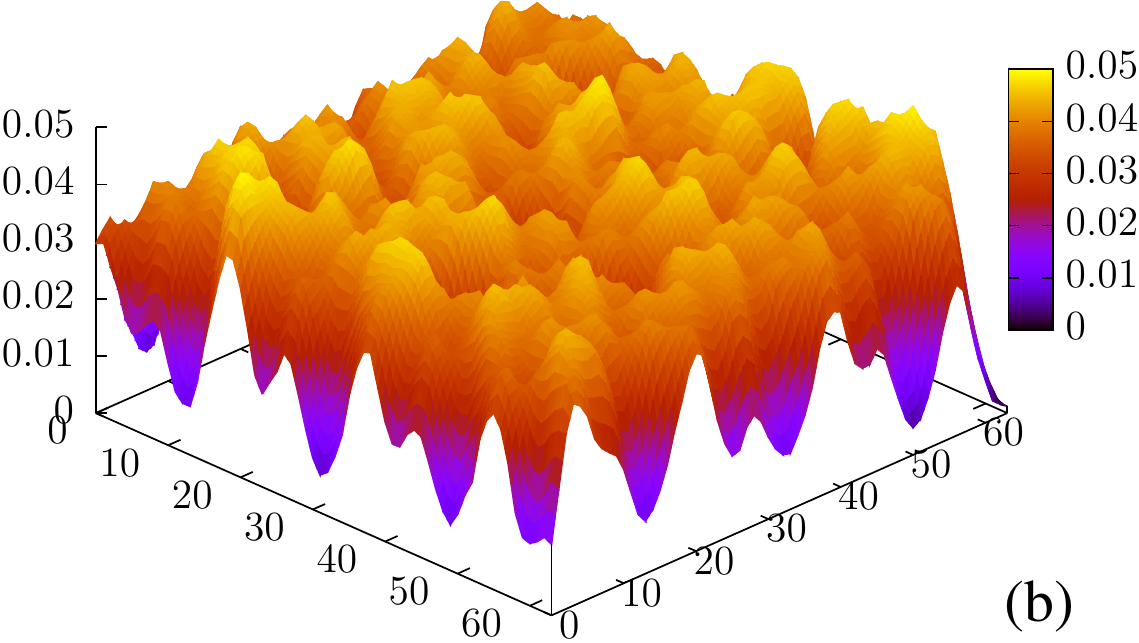}
  \caption{Typical synthetic topography data: (a) top view and (b)  bird's eye view. In (a), the true centers of peaks $\hat{\bm{r}}$ are also represented by circles.} 
  \label{fig:2}
\end{figure}

The amplitude $\{\hat{x}_k\}\text{ ($k = 1, \dots, N_{\text{peak}}$)}$ of a peak is set as a Gaussian random variable with mean 1 and variance $\sigma_{x}$. Using the set of parameters $\{\hat{\bm{x}},\hat{\bm{r}}\}$, the synthetic data $\bm{y}(\hat{\bm{x}},\hat{\bm{r}})$ is generated through the measurement matrix by Eq.~(\ref{eqn:Dmodel}). We fix the following parameters: $\sigma = \sigma_{\text{true}} \equiv 2.25$, $\ell = 64 \text{px}$, $\rho_{\text{vac.}} = 0.02$, $R_{\text{center}}=0.15 \text{px}$, $\sigma_{x} = 0.01$, and $\sigma_{\epsilon} = \num{5d-4}$. The synthetic data used in this section is shown in Fig.~\ref{fig:2}. 

\begin{figure}[htb]
  \includegraphics[width=.5175\linewidth,clip]{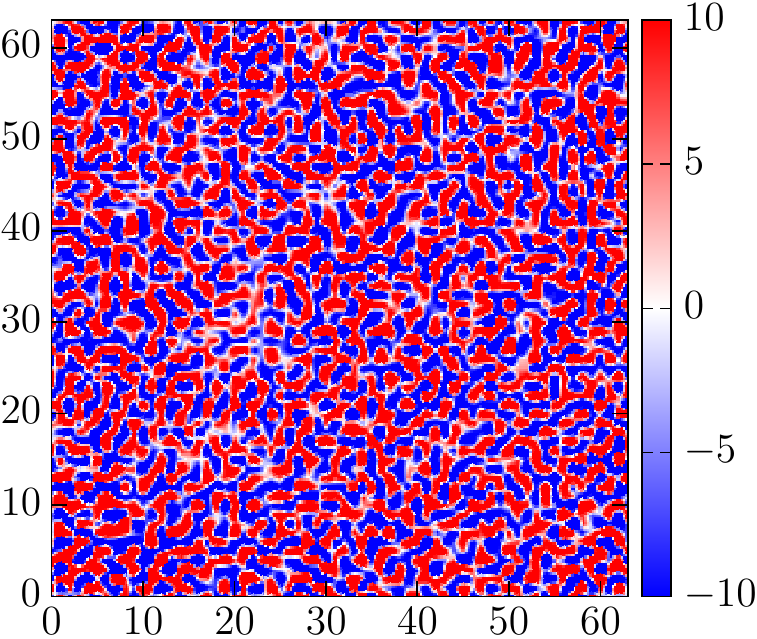}
  \includegraphics[width=.4675\linewidth,clip]{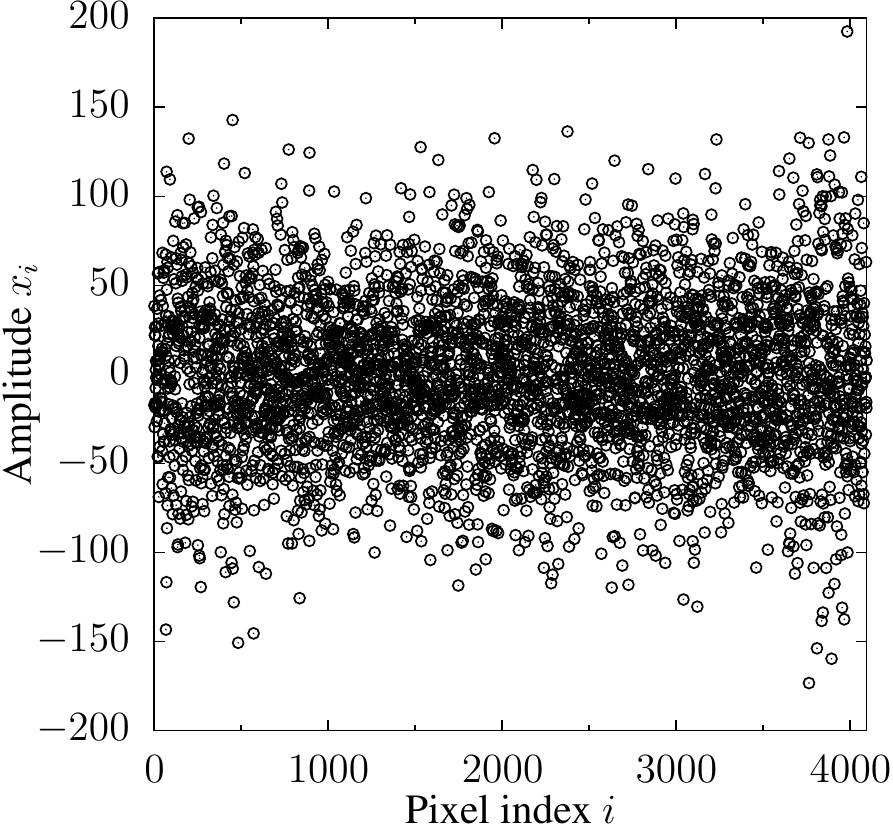}
  \caption{Typical result of the estimated $\bm{x}$ using the least-squares method with $\lambda=0$ and $\sigma=\sigma_{\text{true}}=2.25$. The estimated amplitude $\bm{x}$ is shown in the density plot on the L1VM grid (left) and as a function of the pixel index (right).}
  \label{fig:3}
\end{figure}

For this synthetic data, we first perform the optimization using the
least-squares method ($\lambda = 0$) for fixed $\sigma =
\sigma_{\text{true}}$ ($= 2.25$). As shown in Fig.~\ref{fig:3}, the
optimized vector $\bm{x}$ contains both positive and negative values and
extensively fluctuates with a huge amplitude ($x_i\approx 100$) compared
with the original signal's amplitude ($y_d \approx 0.01$). This result
demonstrates that the least-squares method overfits the data $\bm{y}$,
and a non-sparse solution of $\bm{x}$ is obtained when any regularization terms are absent. 

\begin{figure}[htb]
  \includegraphics[height=3.9truecm,clip]{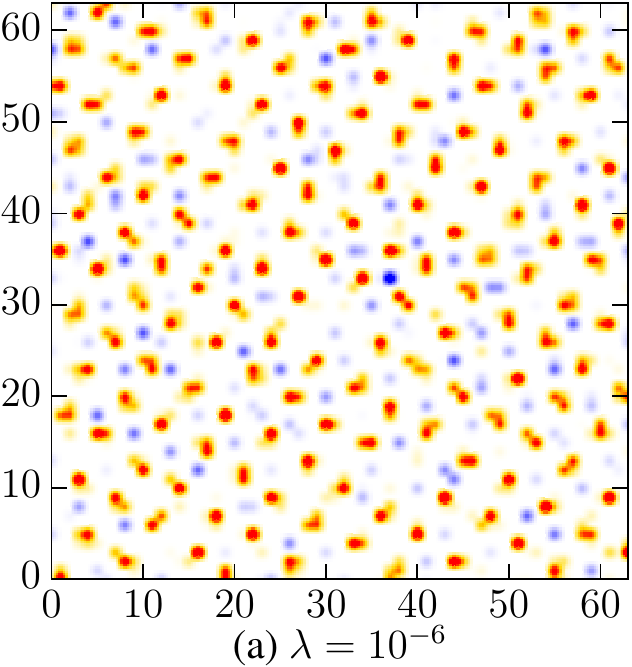}
  \includegraphics[height=4truecm,clip]{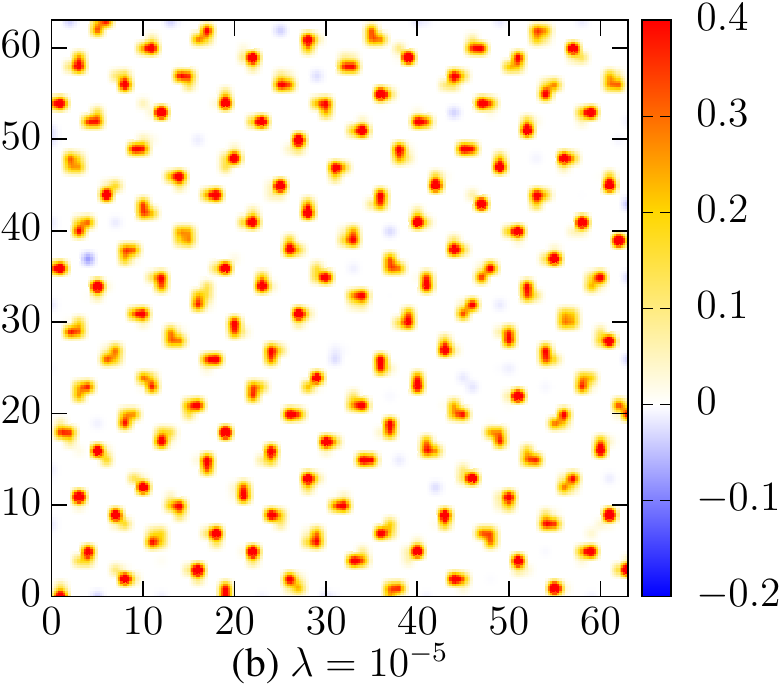}\\
  \includegraphics[height=3.9truecm,clip]{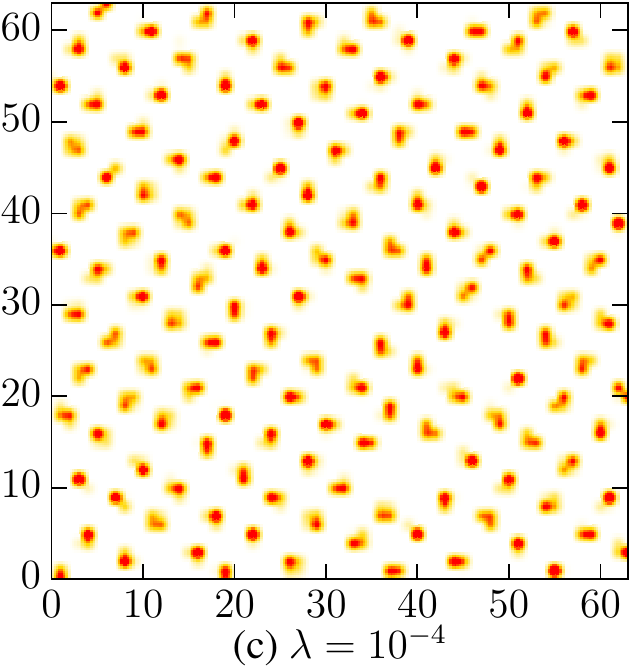}
  \includegraphics[height=4truecm,clip]{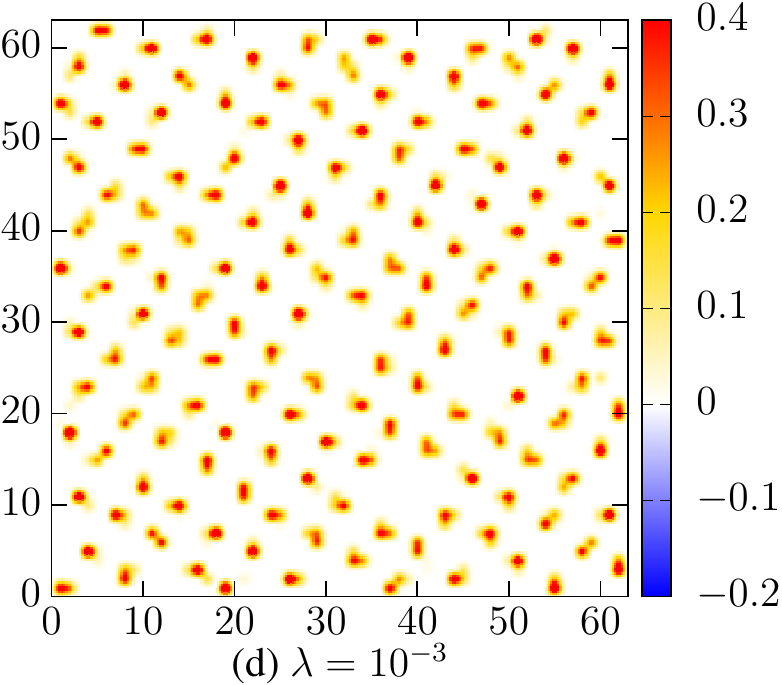}
  \caption{The distributions of typical samples of the optimized $\bm{x}$ with $\sigma=2.25$: (a) $\lambda=10^{-6}$, (b) $\lambda=10^{-5}$, (c) $\lambda=10^{-4}$, and (d) $\lambda=10^{-3}$.}
  \label{fig:4}
\end{figure}

Fig.~\ref{fig:4} shows some typical results of the $\text{L}_1$
optimization with several values of $\lambda$ for fixed
$\sigma=\sigma_{\text{true}}$. While some of the relevant variables have
negative values for a relatively lower value of $\lambda$ such as
$\lambda=10^{-6}$ shown in Fig.~\ref{fig:4}(a), although the true values
$\hat{\bm{x}}$ 
have no negative values. Moreover, the variables are noisy in the
higher-$\lambda$ regime such as $\lambda=10^{-3}$ in comparison with the
$\lambda=10^{-4}$ case. Therefore, there must be a suitable value of
$\lambda$ between these two extremes, which is to be determined using
the CV. 

\subsection{Result of Cross Validation}
\begin{figure}
  \includegraphics[width=0.4925\linewidth,clip]{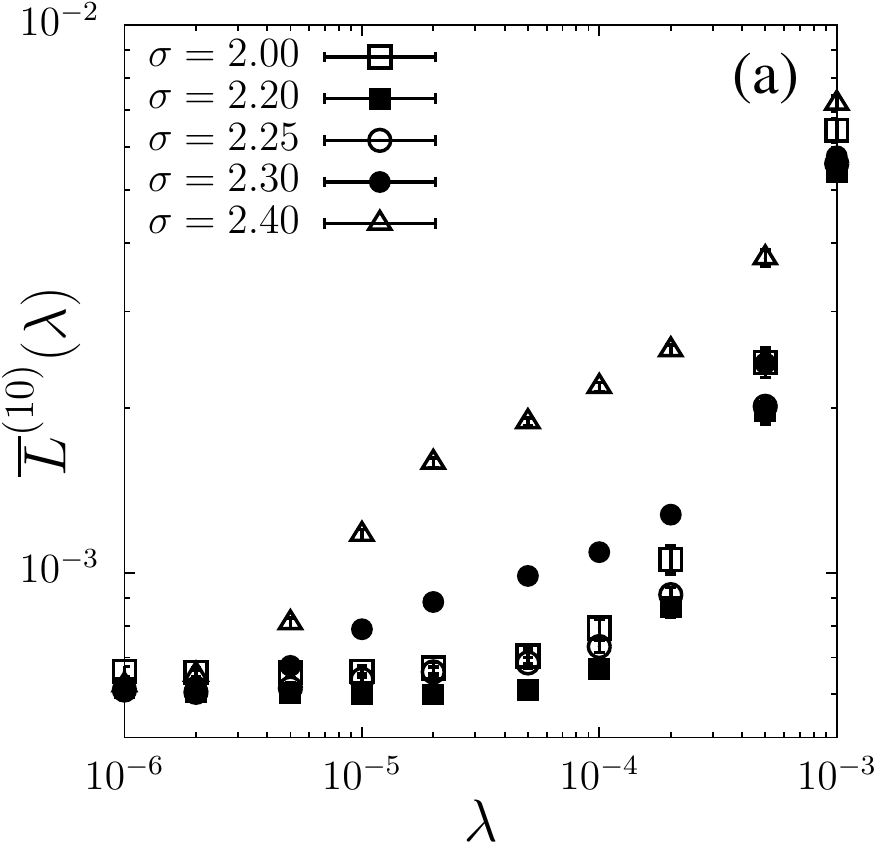}
  \includegraphics[width=0.4925\linewidth,clip]{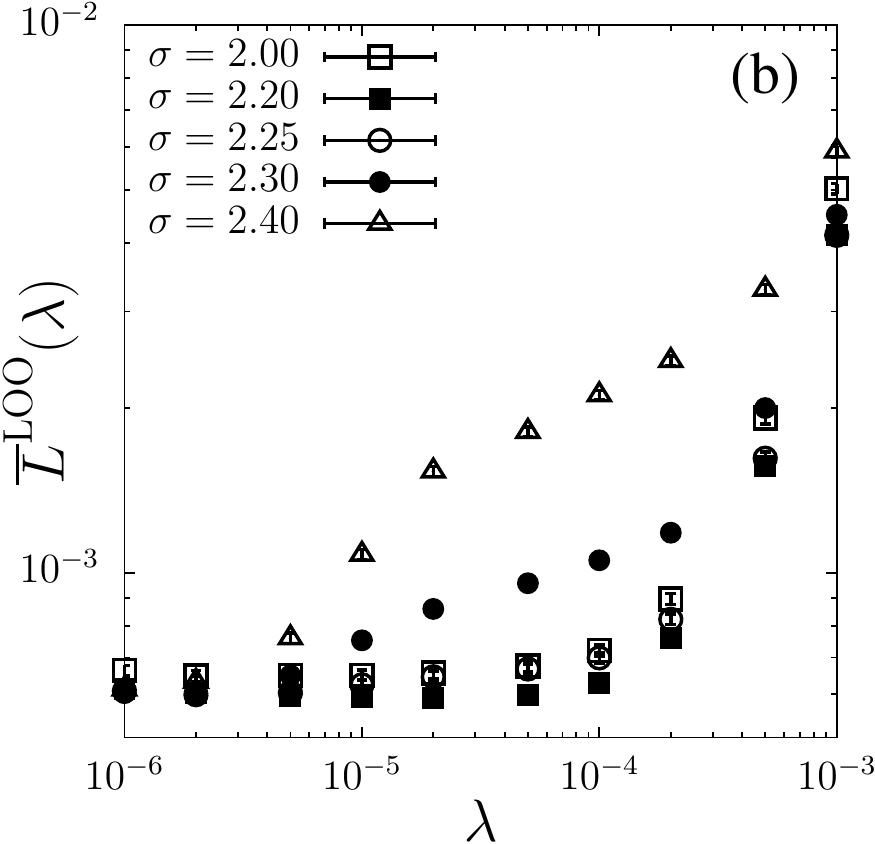}
  \caption{$\lambda$-dependence of the CV errors with different fixed $\sigma$ values for (a) 10-fold CV errors, $\overline{L}^{(10)}(\lambda; \sigma)$ and for (b) the approximated LOOCV errors, $\overline{L}^{\text{LOO}}(\lambda; \sigma)$.}
  \label{fig:5}
\end{figure}
For the synthetic data, we performed the 10-fold CV and LOOCV in order to determine the suitable parameters $\lambda^*$ and $\sigma^*$. The results of 10-fold CV and LOOCV are shown in Fig.~\ref{fig:5}~(a) and (b), respectively. There are no apparent quantitative differences between these results in the regime $10^{-6} < \lambda < 10^{-3}$, indicating that the approximated LOOCV error provides a good estimator of the large $K$-fold CV errors. Then, we choose the optimal $\lambda^*(\sigma)$ for each $\sigma$ in accordance with the one standard error rule. 

\begin{figure}
  \includegraphics[width=\linewidth, clip]{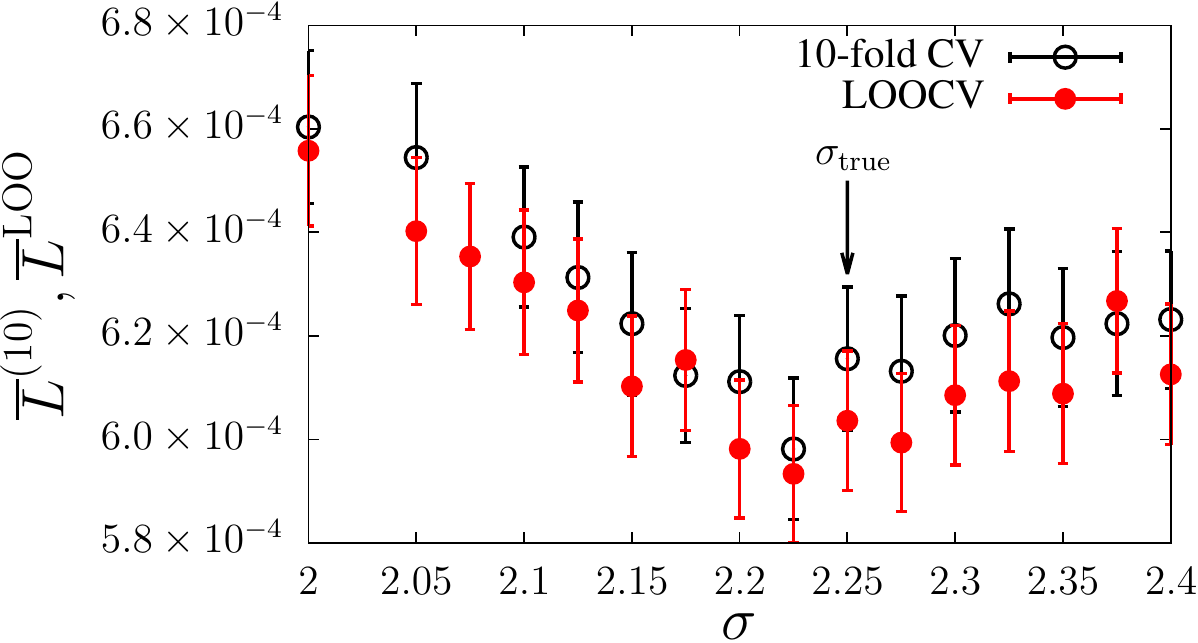}
  \caption{The $\sigma$-dependence of CV errors of 10-fold CV (black open circles) and the approximated LOOCV (red closed circles) for the optimal $\lambda^*(\sigma)$. The arrow indicates the true value of $\sigma$ of the synthetic data.}
  \label{fig:6}
\end{figure}
Next, we study the $\sigma$-dependence of the 10-fold CV error $\overline{L}^{(10)}(\sigma, \lambda^*(\sigma))$ and the LOOCV error $\overline{L}^{\text{LOO}}(\sigma, \lambda^*(\sigma))$ shown in Fig.~\ref{fig:6}. Both CV errors take a minimal value at around $\sigma^* = 2.225$. The error bars displayed in Fig.~\ref{fig:6} represent the standard error of each CV error.  The mean value $\overline{L}$ in the parameter regime $\sigma = 2.225 \pm 0.050$ is within its one standard error at the minimum of $\sigma^* = 2.225$. Hence, this result is consistent with the true value $\sigma_{\text{true}}=2.25$. 

\begin{figure}[htb]
  \includegraphics[width=.6\linewidth,clip]{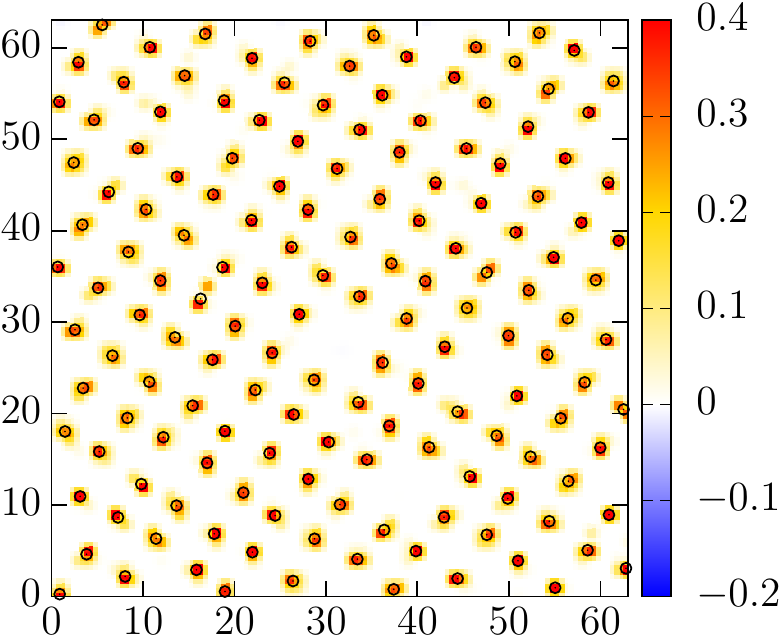}
  \caption{The result of the distribution of $\bm{x}^*$ with $\sigma^*=2.225$ and $\lambda^*(\sigma^*) = \num{2D-5}$. The true peak positions $\hat{\bm{r}}$ are represented by black circles.}
 \label{fig:7}
\end{figure}

By choosing the (hyper)parameters using CV, the optimized amplitude
$\bm{x}^*(\lambda^*; \sigma^*)$ is obtained with $\lambda^* =
\num{2D-5}$ and $\sigma^*=2.225$, which is shown in
Fig.~\ref{fig:7}. The peaks are separated from each other. Thus, we are
able to count the number of the peaks and find $N_{\text{peak}}=153$ in
this case, which is consistent with the number in the synthesis data.

\subsection{Result of estimated atom position}
\begin{figure}[htb]
  \includegraphics[width=.6\linewidth,clip]{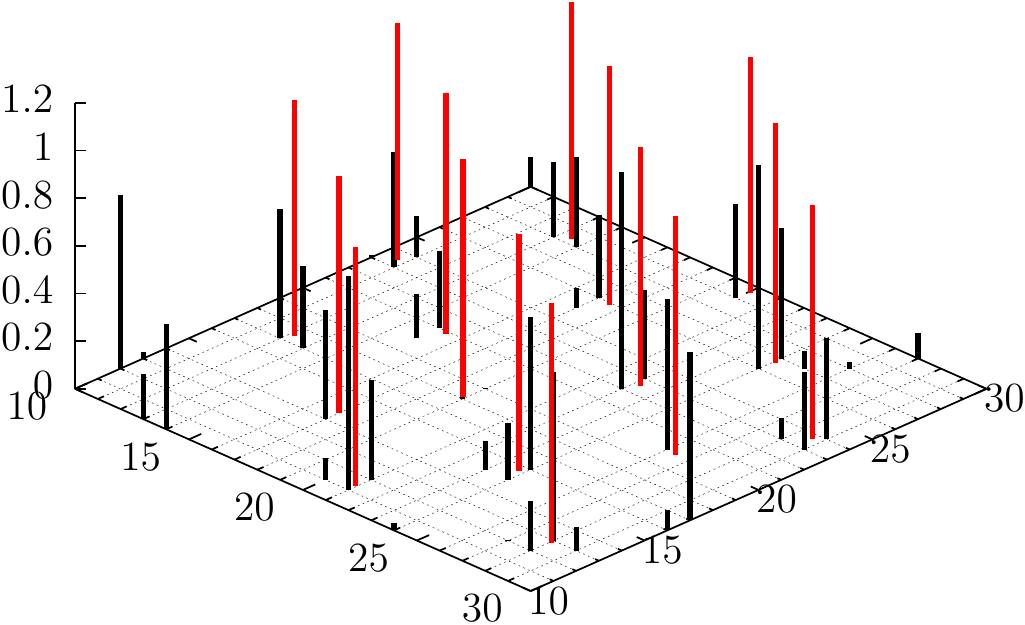}
  \caption{An impulse representation of the estimated amplitude $\bm{x}^*(\sigma^*, \lambda^*(\sigma^*))$ (black) and the true amplitude $\hat{\bm{x}}$ (red).}
  \label{fig:8}
\end{figure}

The final solution $\bm{x}^*(\sigma^*, \lambda^*)$ is sparse, and $x_i$ has a finite value only near the true peak position, as shown in Fig.~\ref{fig:8}. In Fig.~\ref{fig:8}, we show the amplitude of the true peaks $\hat{\bm{x}}$ and the estimated amplitude of L1VM for a portion of the grid . For each true peak, there are still several ``active'' pixels with non-zero elements of $x_i$. The number of active pixels is in the range between two to five, depending on the resolution of the L1VM grid.

\begin{figure}[htb]
  \includegraphics[width=.485\linewidth,clip]{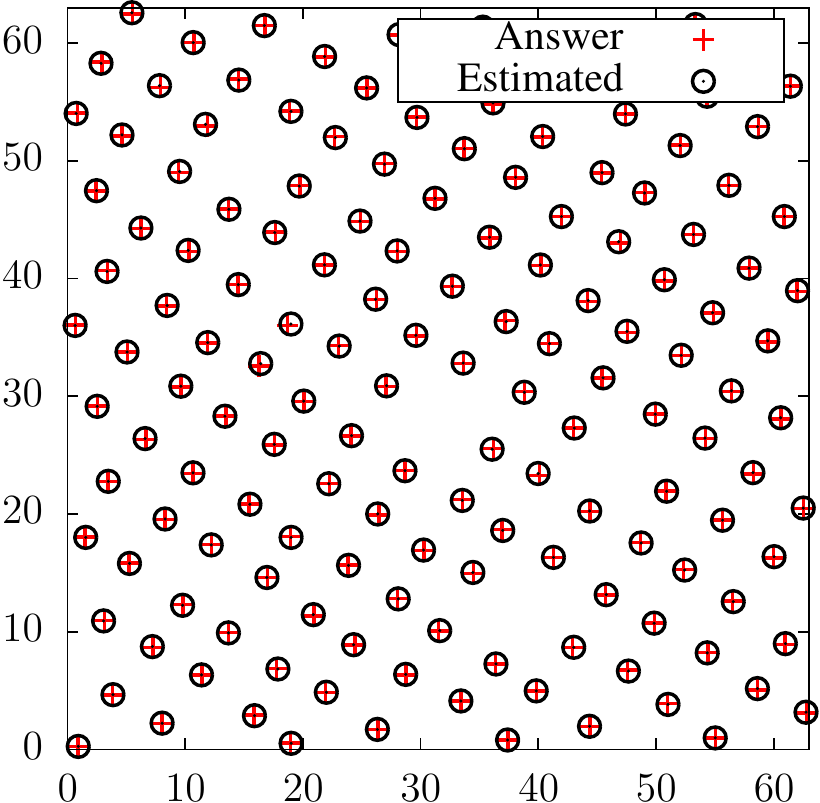}
  \includegraphics[width=.5\linewidth,clip]{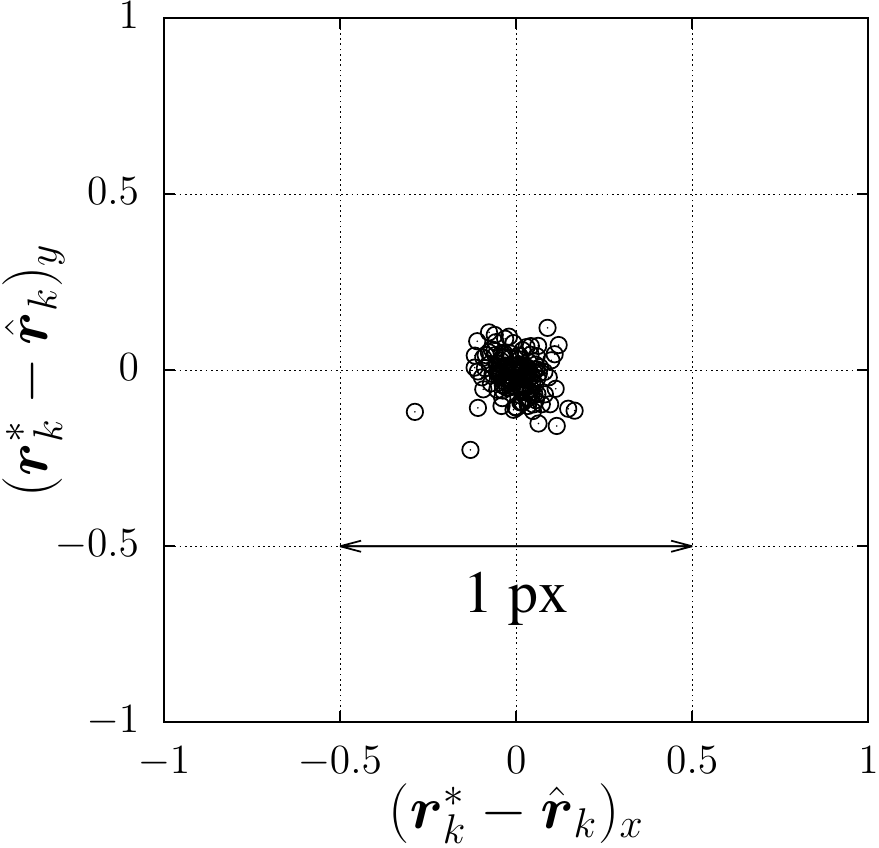}
  \caption{The left view presents peak positions estimated by the $k$-means clustering (white circles) and the true positions (red crosses). The right view shows differences between the true and estimated positions in the $x$ and $y$ directions for each peak.} 
  \label{fig:9}
\end{figure}

Then, we obtained the peak position $\hat{\bm{r}}^*$ by applying the above-mentioned $k$-means clustering method to the optimized L1VM solution $\bm{x}^*(\sigma^*, \lambda^*)$. In Fig.~\ref{fig:9}, we show the obtained positions $\hat{\bm{r}}^*$ together with the true positions $\hat{\bm{r}}$. We also show the difference between the true positions and their corresponding estimated positions on the right side of Fig.~\ref{fig:9}. No significant differences are observed in the figures. In fact, the accuracy of our estimation is within $1$px, meaning that the positions of the peaks are extracted from the STM data with accuracy beyond the resolution of the input signal. This is our main claim in this paper.

\subsection{Application to real experimental data}
\begin{figure*}[t]
  \includegraphics[width=.3\linewidth,clip]{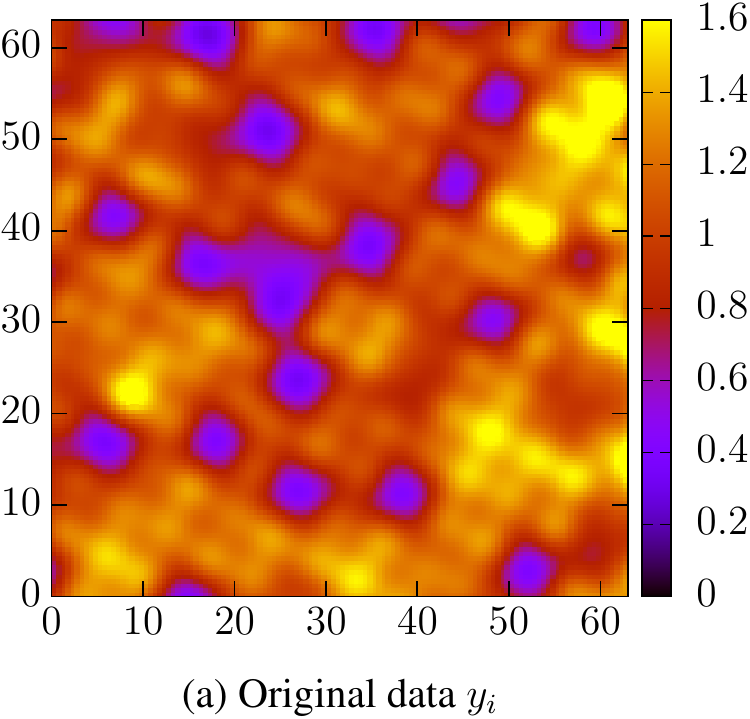}
  \includegraphics[width=.295\linewidth,clip]{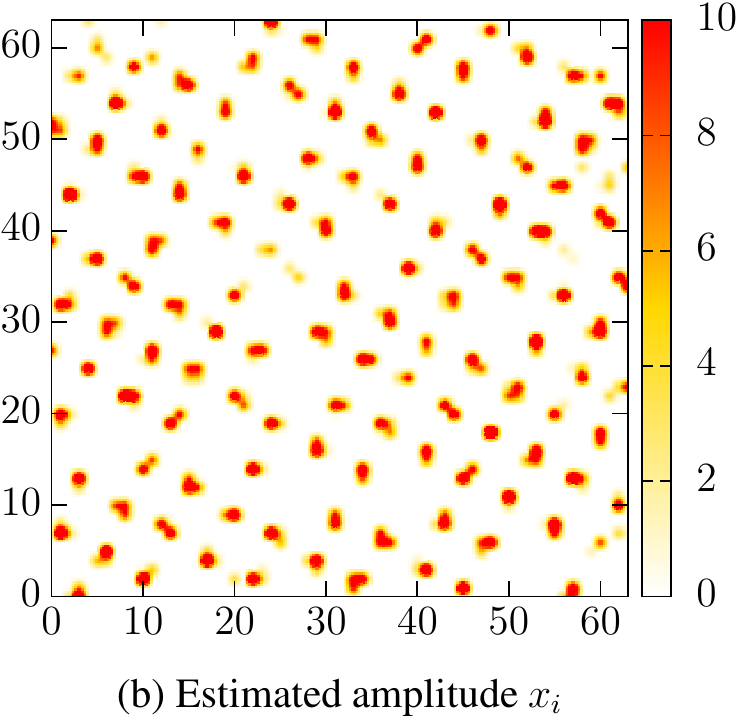}
  \includegraphics[width=.3\linewidth,clip]{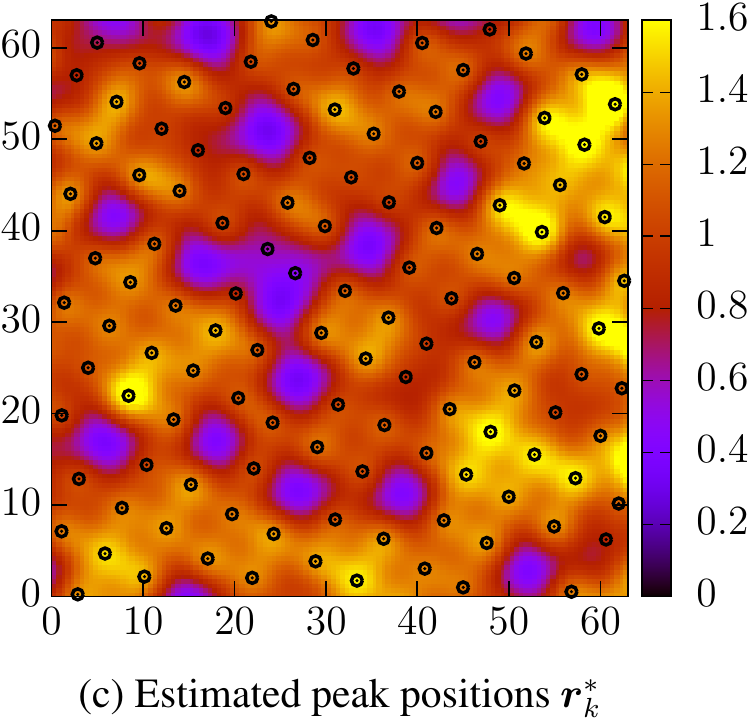}
  \caption{(a) Original topography data obtained by STM measurements on
 a $\text{SrVO}_3$ thin film, provided by Y.~Okada and T.~Hitosugi. The size of data is $64 \times 64$ pixels. (b) Estimated elements of the relevance vector $\bm{x}^*$. (c) Estimated peak positions $\bm{r}^*_k$ obtained by our method on the original topography data.}
  \label{fig:10}
\end{figure*}
The presented results for the synthetic data are useful for examining
the validity of our method. Before applying our scheme to real
experimental data sets, some issues must be addressed. For example, the
choice of the basis function is the one of the essential problems
because the basis function must depend on the surface
materials. However, assuming a Gaussian base function, we apply our
scheme to experimental data from STM topography measurements of a
$\text{SrVO}_3$ thin film. Fig.~\ref{fig:10} presents the tentative
results obtained by our scheme. Many defects are clearly observed on the
square lattice, and the local lattice distortion is enhanced around the
defects. Since our method is not based on Fourier transformations, it
should be possible to directly detect real-space properties such as
local distortion and/or strain.
Details of physical properties of the material are discussed in a
separated paper.

\section{Concluding Remarks}
In this study, we propose an efficient data analysis method for STM topography datasets, which allows highly accurate extraction of peak centers.  Technically, our main problem belongs to a 2D peak decomposition problem with a large unspecified number of peaks . Examples of such problems include NMR spectral data and X-ray or neutron beam diffraction pattern data. Therefore, our scheme could be applicable to a wide range of datasets by changing the basis function.  

First, we discuss the computational cost of our method. An elementary
step of the $\text{L}_1$ optimization consists of 
FISTA. For estimating an $N$ dimensional vector $\bm{x}^*$, the computational cost of FISTA is $O(N^2)$ due to matrix-vector product. The typical computation time required for convergence of $\bm{x}^*(\sigma, \lambda)$ estimation in the analysis of $64 \times 64$ pixel data is about 30 sec with a standard single-core laptop computer. In this case, the dimensions of the measurement matrix $\hat{A}$ is $4096 \times 4096$. The typical size of STM topography data is $512 \times 512$ pixels, so the measurement matrix becomes tremendously large. However, a suitable cutoff length decreases the relevant elements in the measurement matrix when the basis function is spatially localized, such as the Gaussian base function used in this study. We succeeded in a preliminary analysis of $512 \times 512$ pixels of real data using a set of cluster machines.

In our method, most of the computational time is devoted to the hyperparameter estimation by cross validation. As shown in Fig.~\ref{fig:5} and \ref{fig:6}, our results indicate that the approximated LOOCV error proposed by Obuchi and Kabashima  agrees well with the results of the 10-fold CV error. Thus, using the approximated LOOCV, which requires 10 times less computational time than 10-fold CV, is computationally efficient. 

\begin{figure}[ht]
  \includegraphics[width=.9\linewidth,clip]{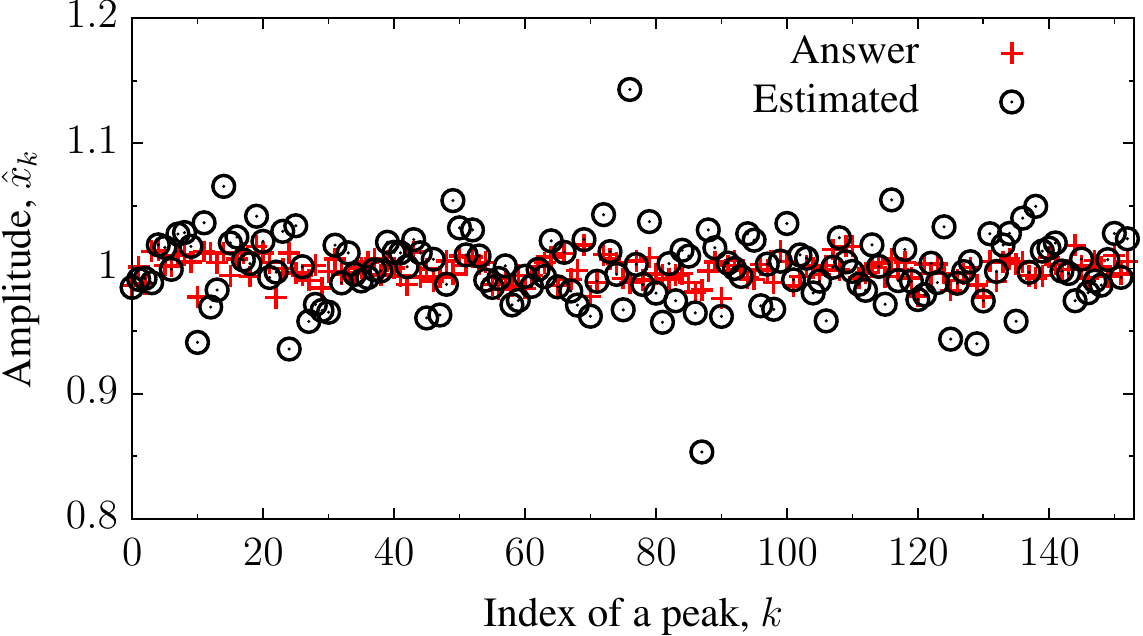}
  \caption{The amplitude of each peak for the estimate $\bm{x}^*$ is shown with an open circle, and the true value $\hat{\bm{x}}$ is shown with a red cross.}
  \label{fig:11}
\end{figure}
\begin{table}[ht]
  \begin{ruledtabular}
    \begin{tabular}{lcc}
      {} & Mean & Standard variance \\
      True $\hat{\bm{x}}$ & $\num{1.00d-3}$ & $\num{9.65d-3}$ \\
      Estimated $\hat{\bm{x}}^*$ & $\num{9.99d-4}$ & $\num{2.96d-2}$
    \end{tabular}
  \end{ruledtabular}
  \caption{The statistical values of the true value $\hat{\bm{x}}$ and the estimate $\hat{\bm{x}}^*$.}
  \label{tab:1}
\end{table}

When we estimate the center positions of the peaks from topography data,
we simultaneously obtain the amplitude values $\bm{x}^*$ of the L1VM
variables. As shown 
in Fig.~\ref{fig:8}, however, our analysis  provides a bundle of peaks for each true peak. In our analysis, the summation of the peak amplitude for each cluster is easily calculated from the optimized variables $\bm{x}^*$ using the attributed variable $z_i$ as
\begin{equation}
  \hat{x}^*_k = \sum_{i=1}^N \delta_{k,z_i} \theta(x_i) x^*_i. 
\end{equation}
In Fig.~\ref{fig:11}, we compare the accumulated amplitude for each peak to the true amplitude. The estimation of the peak amplitude is not accurate, unlike the estimation of the peak position. Our method is significantly modified from the naive least-square method shown in the right side of Fig.~\ref{fig:3}. The mean values and standard variance of the peak amplitudes are shown for our estimate and the true values in Table.~\ref{tab:1}. The mean value of the estimated amplitude $\hat{\bm{x}}^*$ is compatible to that of the true value $\hat{\bm{x}}$, but the standard variance of $\hat{\bm{x}}^*$ is about three times larger than that of $\hat{\bm{x}}$. This discrepancy may be due to the lack of resolution of the L1VM. We expect that the accuracy of the amplitude estimation will be improved by increasing the dimension $N$ of $\bm{x}$ so that $N$ is larger than the input dimension $D$. Another practical way for improving the accuracy might be to re-evaluate the peak amplitude using the knowledge of the peak positions extracted by our scheme.

Finally, the choice of the basis 
function is still an important problem in analyzing experimental
datasets. In the preliminary results shown in Fig.~\ref{fig:10}, typical
STM topography data of a SrVO$_3$ thin film is analyzed by a 2D 
isotropic Gaussian function. However, situations exist where this basis
function choice is not suitable. To apply our scheme to more general
cases, we will utilize machine learning techniques to estimate suitable
basis 
functions from obtained datasets. 

 \acknowledgements{We are grateful to Y.~Okada and T.~Hitosugi for providing STM data and useful discussions. We also thank M.~Okada, Y.~Kabashima, M.~Ohzeki, T.~Obuchi and Y.~Nakanishi-Ohno for useful discussions. This research was supported by the Grants-in-Aid for Scientific Research from the {JSPS}, Japan (No. 25120010  and 25610102) and the Grant-in-Aid for Challenging Exploratory Research from the {MEXT}, Japan (No. 15596332). This work was also supported by ``Materials research by Information Integration'' Initiative (MI$^2$I) project of the Support Program for Starting Up Innovation Hub from Japan Science and Technology Agency (JST).}

\appendix
\section{Fast iterative shrinkage-thresholding algorithm (FITSA)}
\label{sec:FISTA}
In our study, we use FISTA to $\text{L}_1$ minimize the cost function $E(\bm{x})=\lVert \bm{y}-\hat{A}\bm{x}\rVert_2^2/2+\lambda |\bm{x}|_1$. The optimal solution $\bm{x}^*$ can be determined by FISTA by solving iterative equations with auxiliary variable $\beta$ and vector $\bm{\omega}$. One characteristic feature of the algorithm is its use of a soft-thresholding function in the iterative procedure, which is defined by
\begin{equation}
  S_{\epsilon}(x)=
  \begin{cases}
    x - \epsilon & \text{($x > \epsilon$),} \\
    0 & \text{($-\epsilon \leq x \leq \epsilon$),} \\
    x + \epsilon & \text{($x < - \epsilon$).}
  \end{cases}
  \label{eq:SoftThresholding}
\end{equation}
with threshold $\epsilon$. 

By setting the initial conditions $\beta_0=1$ and $\bm{w}_0=\bm{x}_0$, the update procedure is given by the following equations: 
\begin{align}
  \bm{x}_{t+1} &= S_{\lambda/L}(\bm{w}_t
           + \hat{A}^{\text{T}} (\bm{y} - \hat{A} \bm{w}_t) / L), \\
  \beta_{t+1} &= \frac{1+\sqrt{1+4 \beta_t^2}}{2}, \\
  \bm{w}_{t+1} &= \bm{x}_t + \frac{\beta_t - 1}{\beta_{t+1}}
                 (\bm{x}_{t+1} - \bm{x}_t),
\end{align}
where $L$ is a Lipschitz constant of the differential of the squared error, $f(\bm{x})=\lVert\bm{y}-\hat{A}\bm{x} \rVert_2^2/2$, that is, $L$ is a positive constant that satisfies the condition $\lVert\nabla f(\bm{x})-\nabla f(\bm{y})\rVert_2\leq L \lVert\bm{x}-\bm{y}\rVert_2$. Thus, it is natural to choose the threshold $\epsilon_{\text{th}}$ as $\epsilon_{\text{th}}=\lambda/L$ in Eq.~(\ref{eq:LOOCV}).

The Lipschitz constant $L$ is given as $L=\lVert\hat{A}^{\text{T}}\hat{A}\rVert$ with $\lVert\cdots\rVert$ being an operator norm of a matrix. The value of $L$ is computable when the matrix is smaller than a $100 \times 100$ matrix. For a larger matrix, $L$ can be estimated using the backtracking algorithm \cite{FISTA}. Moreover, the sum of a column of our measurement matrix $\hat{A}$ takes a value close to unity, yielding $\lVert\hat{A}^{\text{T}}\hat{A}\rVert\approx 1$ for a large matrix $\hat{A}$ by a simple calculation of linear algebra.

\end{document}